# Second harmonic generation in suspensions of spherical particles

E.V. Makeev[1] and S.E. Skipetrov[2]

[1] Physics Department, Moscow State University, 119992 Moscow, Russia

[2] CNRS/Laboratoire de Physique et Modélisation des Milieux Condensés, 25 Avenue des Martyrs, 38042 Grenoble, France

**Abstract**

We study the second harmonic generation (SHG) in a suspension of small spherical particles confined within a slab, assuming undepleted pump and applying (i) single scattering approximation and (ii) diffusion approximation. In the case (i), the angular diagram, the differential and total crossections of the SHG process, as well as the average cosine of SH scattering angle are calculated. In the case (ii), the average SH intensity is found to show no explicit dependence on the linear scattering properties of the suspension. The average intensity of SH wave scales as $I_0 L/\Lambda_2$ in both cases (i) and (ii), where $I_0$ is the intensity of the incident wave, $L$ is the slab thickness, and $\Lambda_2$ is an intensity-dependent "SH scattering" length.

---

[1] E-mail: makeev@newmail.ru
[2] Corresponding author. E-mail: Sergey.Skipetrov@grenoble.cnrs.fr

# 1. Introduction

Nonlinear optics of disordered media is a rapidly developing branch of modern optics [1,2]. The interplay of disorder and nonlinearity can lead to qualitatively different results depending on their relative strength. A relatively well studied case is that of laser beam propagation in a nonlinear medium with an extremely weak disorder (e.g., turbulent atmosphere), where the transport mean free path $l^*$ due to scattering is much larger that the characteristic length $l_{\mathrm{NL}}$ of nonlinear interaction: $l^* \gg l_{\mathrm{NL}}$ (see Ref. 2 for a review). In this case, the nonlinear phenomena develop before the scattering destroys the initial beam and thus can be studied in the framework of paraxial (or parabolic) approximation, providing an adequate description of light propagation at short distances $z$ from the source of radiation ($z \ll l^*$ but $z > l_{\mathrm{NL}}$). Paraxial approximation consists in assuming the variation of the amplitude $\vec{A}(\vec{r} = \{\vec{\rho}, z\})$ of electric field $\vec{E}(\vec{r},t) = \vec{A}(\vec{r})\exp(ikz - i\omega t)$ of the beam along the direction $z$ of beam propagation to occur only over distances much larger than the optical wavelength, and neglecting the second-order derivative $\partial^2 \vec{A}/\partial z^2$ in the wave equation (see, e.g. Ref. 3, p. 91). This approximation is justifiable insofar as the wave $\vec{E}(\vec{r},t)$ is propagating primarily along the $z$ axis (which is true for $z \ll l^*$). In this regime, propagation of light is governed by nonlinear effects, while scattering can be taken into account as a small perturbation. Despite a considerable simplification of the problem due to the replacement of the full nonlinear wave equation by a paraxial one, this type of problems is still under active study (e.g., in connection with the filamentation of powerful laser pulses in the atmosphere [4,5]).

From the theoretical standpoint, a much more complicated problem is that of light propagation in strongly disordered nonlinear media, where $l^* \ll l_{\mathrm{NL}}$ and $z > l^*$. Paraxial approximation does not apply in this case since the initial laser beam is completely destroyed by scattering well before the nonlinear phenomena start to play a role. One then deals with nonlinear interaction of diffuse waves. This problem has been studied in nonlinear media with cubic nonlinearity (Kerr effect) and proved to be a nontrivial one [6,7]. On the other hand, the lowest-order nonlinear effects – those quadratic in the wave field – have also been considered in the limit $l^* \ll l_{\mathrm{NL}}$ [8]. Recent theoretical [9,10] and experimental [11] work suggests that even the simplest realization of a nonlinear disordered medium, a dilute suspension of spherical particles, exhibits a rather unexpected and nontrivial behavior: second harmonic (SH) is generated by a single spherical particle despite its central symmetry and seemingly prohibited second-order nonlinear effects, and a measurable SH signal can be obtained from a dilute suspension of such particles in a single scattering regime [11]. Note that although SHG in 1D disordered systems of spheres has been studied theoretically in Refs. 12, the results of that study cannot be applied to describe SHG in 3D samples since disorder is known to have qualitatively different impacts on wave propagation in systems of different dimensionalities (for example, there is no diffusive regime in 1D) [13].

In the present paper, we provide a theoretical description of second harmonic generation (SHG) in a suspension of small spherical particles, restricting ourselves to the following limiting cases: (i) dilute suspension (sample size $L \ll$ than the scattering length $l \leq l^*$) and (ii) concentrated suspension ($L \gg l^*$). In the former case, we apply a single scattering approximation, while in the latter case a diffusion approximation for both the fundamental and SH fields seems to be adequate. While the results obtained for the case (i) are in qualitative agreement with experimental data of Ref. 11 (quantitative agreement is not expected since the experiments of Ref. 11 were performed with particles of the order or larger than optical wavelength), new theoretical predictions are made for the case (ii), calling for new experiments.

## 2. SHG by an isolated spherical particle and a dilute suspension of particles

We start with a brief derivation of the main equations describing SHG by a small isolated spherical particle of radius $R \ll \lambda$ ($\lambda$ is the wavelength of the incident wave) surrounded by the free space (see Fig. 1). Free and bounded electric charges of the particle (total charge remaining zero) redistribute in space under the influence of electric field of the incident wave $\vec{E}_0(\vec{r})\exp(-i\omega t)$, the multipole (dipole, quadrupole, etc.) moments of this new charge distribution being nonzero [9,10]. These multipole moments oscillate in time due to the oscillations of the incident electric field and hence the particle re-radiates electromagnetic waves. We will be interested in the re-radiation at the fundamental frequency $\omega$ ("linear" scattered wave) and doubled frequency $2\omega$ (SHG or "SH scattering" for brevity). While the linear scattering process is well described by the standard scattering theory [14], the SH scattering requires special attention. Because of the central symmetry of the particle, the SH dipole moment of the particle $\vec{p}\exp(-2i\omega t)$ has no "local" contribution proportional to $\vec{E}_0\vec{E}_0$. However, due to the spatial heterogeneity of the field $\vec{E}_0(\vec{r})$, there exists a "nonlocal" contribution to the SH dipole moment: $\vec{p} \propto \vec{E}_0 \nabla \vec{E}_0$, where the values of $\vec{E}_0$ and $\nabla \vec{E}_0$ are taken in the center of the particle and the *i*-th component of $\vec{E}_0 \nabla \vec{E}_0$ is defined as $E_{0j}(\partial E_{0i}/\partial x_j)$. From here on we assume summation over repeated indices (*j* in the present case). By virtue of the spherical symmetry, $\vec{p}$ can be written as [9]

$$\vec{p} = \gamma^d (\vec{E}_0 \nabla \vec{E}_0), \tag{1}$$

where $\gamma^d$ is a scalar representing the dipole nonlinear polarizability of the particle. The leading contribution to the SH quadrupole moment is, in contrast, of local nature [9]:

$$Q_{ij} = \gamma^Q \left[ E_{0i} E_{0j} - \frac{1}{3} \delta_{ij} E_0^2 \right], \tag{2}$$

where the scalar quantity $\gamma^Q$ is the quadrupole nonlinear polarizability. Consequently, the electric field radiated by the particle at frequency $2\omega$ is

$$\vec{E}_2(\vec{r}) = k_2^2 \frac{\exp(ik_2 r)}{r} (\hat{n} \times \vec{p}^{\,eff}) \times \hat{n}, \tag{3}$$

where $\hat{n}$ is a unit vector in the direction of observation, $\vec{k}_2$ is the wave vector of the SH wave outside the particle, and $\vec{p}^{\,eff}$ is the effective dipole moment:

$$p_i^{eff} = (p_i - \frac{i}{6} k_2 Q_{ij} \hat{n}_j) = \left[ \gamma^d E_{0j} (\partial E_{0i}/\partial x_j) - \frac{i}{6} k_2 \gamma^Q (E_{0i} E_{0j} \hat{n}_j - \frac{1}{3} E_{0k} E_{0k} \hat{n}_i) \right]. \tag{4}$$

Note that Eq. (4) contains two distinct contributions: the dipole one and the quadrupole one. Depending on the values of $\gamma^d$ and $\gamma^Q$, they can be either comparable or one of the contributions can dominate the other.

The power of SH radiated by the spherical particle in a unit solid angle follows directly from Eq. (3):

$$\frac{dP_2}{d\Omega} = \frac{c}{8\pi} k_2^4 \left| (\hat{n} \times \vec{p}^{\,eff}) \times \hat{n} \right|^2. \tag{5}$$

The differential crossection of SHG process is then

$$\frac{d\Sigma_2}{d\Omega} \equiv \frac{1}{I_0^2}\frac{dP_2}{d\Omega} = \frac{8\pi}{c}k_2^4 \frac{|(\hat{n}\times\vec{p}^{\,eff})\times\hat{n}|^2}{|\vec{E}_0|^4}, \quad (6)$$

where $c$ is the speed of light in the free space and $I_0 = [c/(8\pi)]\cdot|E_0|^2$ is the intensity of the incident wave.

The above expressions are valid, generally speaking, for both linearly and circularly polarized incident fundamental waves. We now consider a particular case of a linearly polarized incident plane wave $\vec{E}_0(\vec{r}) = E_0 \exp(i\vec{k}\vec{r})\hat{\varepsilon}_0$, where $E_0$ is the amplitude of the wave, $\hat{\varepsilon}_0$ is a unit vector in the direction of its polarization, and $\vec{k} = (\omega/c)\hat{k}$ with $\hat{k}$ a unit vector in the direction of propagation. Eqs. (3) and (4) then yield [10]

$$\vec{E}_2 = k_2^2 \frac{\exp(ik_2 r)}{r}[\hat{n}\times(\vec{p} - \frac{ik_2}{6}\vec{Q}(\hat{n}))]\times\hat{n}, \quad (7)$$

where the vector $\vec{Q}(\hat{n})$ is defined by $Q_i(\hat{n}) = Q_{ij}\hat{n}_j$, and [10]

$$\vec{p} = \frac{8\pi i k}{15} R^3 E_0^2 \chi_1 \hat{k}(\hat{\varepsilon}_0 \cdot \hat{\varepsilon}_0), \quad (8)$$

$$\vec{Q}(\hat{n}) = \frac{16\pi}{5} R^3 E_0^2 \chi_2 (\hat{n}\cdot\hat{\varepsilon}_0)\hat{\varepsilon}_0. \quad (9)$$

In the above equations we have introduced new parameters $\chi_1$ and $\chi_2$ related to $\gamma^d$ and $\gamma^Q$ by $\gamma^d = (8\pi/15)\chi_1 R^3$ and $\gamma^Q = (16\pi/5)\chi_2 R^3$. Calculation of $\chi_1$ and $\chi_2$ for a given material can be accomplished using the results of Ref. 10 and is not a subject of the present paper. We note, however, that, in general, $\chi_1$ and $\chi_2$ are complex-valued functions of frequency. Below we consider only the case of purely *real* $\chi_1$ and $\chi_2$, commenting, where appropriate, on how the results are modified in the case of complex $\chi_1$ and $\chi_2$.

According to Eq. (5), the power of SH wave radiated in a unit solid angle becomes [10]

$$\frac{dP_2}{d\Omega} = \frac{ck_2^4}{2\pi} \times$$
$$\times \left\{ |\vec{p}|^2 [1-(\hat{n}\cdot\hat{k})^2] + \left(\frac{k_2}{6}\right)^2 |\vec{Q}(\hat{n})|^2 [1-|\hat{n}\cdot\hat{\varepsilon}_0|^2] + \frac{k_2}{3}\mathrm{Im}[(\hat{n}\cdot\vec{p})(\hat{n}\cdot\vec{Q}(\hat{n}))^*]\right\}. \quad (10)$$

Note that the first term in the curved brackets of Eq. (10) corresponds to SHG due to the nonlocal dipole radiation of the sphere, the second term is due to the local quadrupole radiation, while the third and the last term is a combined term originating from the interference of dipole and quadrupole radiation.

The main results of the above analysis — Eqs. (6) and (10) — have been previously derived in Refs. 9 and 10. We now analyze them in more detail. Choosing the $z$ axis parallel to $\vec{k}$ and the $y$ axis parallel to $\hat{\varepsilon}_0$ as shown in Fig. 1, we put Eq. (10) in the following form:

$$\frac{dP_2}{d\Omega} = P_0 \left\{ \chi_1^2 F_1(\theta,\varphi) + 4\chi_2^2 F_2(\theta,\varphi) + 4\chi_1\chi_2 F_3(\theta,\varphi) \right\}, \quad (11)$$

where $\theta$ is the angle between the $z$ axis and the direction of observation, $\varphi$ is the corresponding polar angle, $P_0 = (512/225)\pi c(kR)^6 E_0^4$, and

$$F_1(\theta,\varphi) = \sin^2\theta, \quad (12)$$
$$F_2(\theta,\varphi) = \sin^2\theta \sin^2\varphi (1-\sin^2\theta\sin^2\varphi), \quad (13)$$
$$F_3(\theta,\varphi) = \cos\theta \sin^2\theta \sin^2\varphi \quad (14)$$

are scalar functions, describing angular diagrams of the dipole radiation, the quadrupole radiation and the interference term, respectively. We show the angular diagrams given by $F_1(\theta, \varphi)$, $F_2(\theta, \varphi)$ and $F_3(\theta, \varphi)$, as well as their sum at $\chi_1 = \chi_2$ in Fig. 2. Note that $F_3 < 0$ in the backward hemisphere (for $\pi/2 < \theta < \pi$) and Fig. 2(c) shows the absolute value of $F_3$. It follows from Eq. (11) that if $|\chi_1| \sim |\chi_2|$, the three terms in Eq. (11) are of the same order, while the dipole (quadrupole) term dominates at $|\chi_1| \gg |\chi_2|$ ($|\chi_1| \ll |\chi_2|$). Eqs. (11)-(14) lead to the following expression for the differential and total crossections of SHG (or SH scattering) by an isolated spherical particle:

$$\frac{d\Sigma_2}{d\Omega} = \frac{1}{I_0^2}\frac{dP_2}{d\Omega} = (8\pi)^3 \frac{64}{225}\frac{(kR)^6}{c}\left\{\chi_1^2 F_1(\theta,\varphi) + 4\chi_2^2 F_2(\theta,\varphi) + 4\chi_1\chi_2 F_3(\theta,\varphi)\right\}, \quad (15)$$

$$\Sigma_2 = \int_{4\pi}\frac{d\Sigma_2}{d\Omega}d\Omega = (8\pi)^4 \frac{64}{675}\frac{(kR)^6}{c}\left\{\chi_1^2 + \frac{4}{5}\chi_2^2\right\}. \quad (16)$$

As one can see from Eq. (16), $\Sigma_2$ is proportional to $(R/\lambda)^6$ in contrast to the crossection $\sigma$ of linear Rayleigh scattering, which is known to scale as $(R/\lambda)^4$ [14]. This difference is qualitatively explained by the fact that the linear scattered wave originates from the dipole radiation with the dipole moment $\propto \vec{E}_0$, while in the case of SH scattering an equivalent, local dipole contribution is prohibited by symmetry and the radiated SH field is due to the second-order contribution to the dipole moment $\vec{p} \propto \vec{E}_0 \nabla \vec{E}_0$ [first term in Eq. (16)] and the quadrupole moment [second term in Eq. (16)]. It is interesting to note that the third, interference term in the angular brackets of Eq. (15) for the differential SH scattering crossection does not contribute to the total crossection (16). Hence, the role interference between dipole and quadrupole radiations amounts to redistribute the SH energy between different directions but not to modify the total radiated energy. In the case of complex $\chi_1$ and/or $\chi_2$, $\chi_1^2$, $\chi_2^2$ and $\chi_1\chi_2$ in Eqs. (11), (15) and (16) should be replaced by $|\chi_1|^2$, $|\chi_2|^2$ and $\mathrm{Re}(\chi_1\chi_2^*)$, respectively. Obviously, this does not change the angular dependence of the three terms contributing to the angular diagram of SHG but modifies their relative weights.

While Eq. (16) describes the total power of SHG by a small spherical particle, the anisotropy of SHG can be characterized by the average cosine of SH scattering angle:

$$\langle\cos\theta\rangle = \frac{1}{\Sigma_2}\int_{4\pi}\frac{d\Sigma_2}{d\Omega}\cos\theta d\Omega = \left(\frac{5}{2}\zeta + \frac{2}{\zeta}\right)^{-1}, \quad (17)$$

which appears to depend on a single parameter $\zeta \equiv \chi_1/\chi_2$. Eq. (17) exhibits a remarkably nonmonotonic behavior (see Fig. 3). While $\langle\cos\theta\rangle$ is small and hence roughly equal power is scattered to the forward and backward hemispheres at $|\zeta| \gg 1$ (dominance of the dipole radiation) and $|\zeta| \ll 1$ (dominance of the quadrupole radiation), it reaches a maximum (minimum) $\langle \cos\theta \rangle_{max} = \pm 1/(2\sqrt{5}) \approx \pm 0.22$ at $\zeta = \pm 2/\sqrt{5} \approx \pm 0.89$, when the interference term comes into play. At $|\chi_1| \sim |\chi_2|$ the SH scattering is hence anisotropic with preference for the forward (if $\chi_1\chi_2 > 0$) or backward (if $\chi_1\chi_2 < 0$) direction [see also Fig. 2(d)]. It is worth noting that while the interference term in Eq. (15) for the differential crossection does not contribute to the total crossection of SH scattering, Eq. (16), (as we have noted above), it is this and only this term that leads to the anisotropy of SH scattering, since the pure dipole and quadrupole terms are symmetric with respect to a mirror transformation $z \to -z$ and in the absence of the interference term would lead to $\langle \cos\theta \rangle = 0$. In the case of complex $\chi_1$ and/or $\chi_2$, $(5/2)\zeta$ and $2/\zeta$ in Eq. (17) should be replaced by $(5/2)|\chi_1|^2/\text{Re}(\chi_1\chi_2^*)$ and $2|\chi_2|^2/\text{Re}(\chi_1\chi_2^*)$, respectively, and the average cosine of SH scattering angle is no longer parameterized by a single parameter $\zeta$.

Taking into account that the SH signal generated by a single spherical particle is, in general, small (partly because $\chi_1$ and $\chi_2$ are small, and partly because $\Sigma_2 \propto (R/\lambda)^6$ and $R/\lambda \ll 1$), its experimental observation can be accomplished for an ensemble (suspension) of many particles only. We therefore consider a slab of surface area *A*, perpendicular to the *z* axis and located in between the planes *z* = 0 and *z* = *L*, filled with a dilute suspension of identical spherical particles (particle number density *n*). We assume that *n* is small enough for the condition *L* << *l*, *l*$_2$ to be fulfilled. Here $l = 1/n\sigma$ and $l_2 = 1/n\sigma_2$ are the linear scattering lengths and $\sigma$ and $\sigma_2$ are the linear scattering crossections for the fundamental wave and the second harmonic, respectively. The SH scattering of an incident plane wave in the suspension can be then treated in the single scattering approximation. In addition, we assume that only a small amount of energy of the incident wave is transferred to the SH wave and that the intensity of the fundamental wave always remains much larger that the intensity of the SH wave. We can then neglect the decrease of the intensity of the fundamental wave as it propagates through the suspension (the so-called undepleted-pump approximation) and assume than all particles in the suspension experience the incidence of the same plane, linearly polarized wave $\vec{E}_0(\vec{r})\exp(-i\omega t)$. The differential and total SH scattering crossections of suspension are then simply given by the corresponding results for a single particle, multiplied by the total number of particles *N* = *nAL*. The average power of SH scattering per unit solid angle becomes

$$\frac{dP_2^{(N)}}{d\Omega} = \frac{d\Sigma_2}{d\Omega} nI_0^2 LA = \frac{d\Sigma_2}{d\Omega} \frac{A}{\Sigma_2} \frac{L}{\Lambda_2} I_0, \qquad (18)$$

while the average intensity of the SH wave at $z > L$ is

$$I_2 = \frac{1}{A} \int_{2\pi} \frac{dP_2^{(N)}}{d\Omega} d\Omega = \frac{1}{2} \frac{L}{\Lambda_2} I_0 \left(1 + \frac{15}{8} \langle \cos\theta \rangle \right), \qquad (19)$$

where we introduce a characteristic "SH scattering length"

$$\Lambda_2 = \frac{1}{\Sigma_2 n I_0}. \qquad (20)$$

It is worthwhile to note that $I_2 \propto L$. The angular diagrams of SH scattering (see Fig. 2) as well as the average cosine of the scattering angle (Fig. 3) coincide with those obtained for an isolated particle [see Eqs. (11), (17)]. A characteristic dip in the forward direction seen in the diagram of Fig. 2(d) is in a good agreement with the recent experimental measurements [11]. A quantitative comparison of the theory and experiment is complicated by the fact that the size of colloidal particles used in Ref. 11 was of the order or even larger than the wavelength $\lambda$. Quantitative description of this case requires more sophisticated theoretical treatment and is not a subject of the present paper.

To conclude this section, we comment on the issue of phase matching known to be very important in the context of optical harmonic generation (see, e.g., Ref. 3, pp. 58 and 65). In the above analysis of SHG in a dilute suspension of small particles, we assume that SH waves generated by individual particles add incoherently at the measurement point (leading to linear scaling of SH intensity with particle number). This assumption is justified as long as (a) SH wave generated by a given particle acquires large random phase shift, uncorrelated with phase shifts of SH waves generated by other particles, during its propagation to the measurement point, and (b) phase difference between SH waves generated by any two (even neighboring) particles is typically $\gg 2\pi$. Both conditions (a) and (b) should be satisfied for a dilute suspension where (a) particle positions are random and completely uncorrelated and (b) typical inter-particle distance $d$ much exceeds the optical wavelength $\lambda$. In the experiments of Ref. 11, for example, $d \sim 10$ $\mu$m which is at least an order of magnitude larger than $\lambda$. This illustrates the relevance of our analysis to real experimental situations. However, if the particle concentration is increased and $d$ approaches $\lambda$, or if some order is introduced in the spatial distribution of particles in suspension, our results will not be valid any longer and one has to take into account the (partially) coherent addition of SH waves generated by different particles ("collective" SH scattering). Finally, relation between dielectric constants $\varepsilon(\omega)$ and $\varepsilon(2\omega)$ inside the particle (phase matching inside the particle) can also play some role but enters our results only via parameters $\chi_1$ and $\chi_2$ (which are sensitive to $\varepsilon(\omega)$ and $\varepsilon(2\omega)$ [10]) without any influence on the qualitative features of SHG discussed above.

## 3. SHG by a concentrated suspension of spherical particles

In order to enhance the SH signal, one could try to increase the density $n$ of particles in the suspension. As $n$ increases, the scattering lengths $l$, $l_2$ decrease, the condition $L \ll l, l_2$ breaks down, and the single scattering approximation adopted in the previous section becomes invalid [15]. While the intermediate regime $L \sim l, l_2$ is generally hard to describe even in the case of pure linear scattering [14,15], both incident and SH wave can be treated in the diffusion approximation [15] in the limit $L \gg l^*, l_2^*$, if $kl^* \gg 1$ and $k_2 l_2^* \gg 1$. Physically, this means that the incident wave at frequency $\omega$ experiences multiple scattering events before leaving the sample, SH being generated at each scattering event. Similarly, the SH waves are generated all over the sample and experience multiple elastic scattering before leaving the sample. Again, as in the previous section, we assume low SH intensity and work in the undepleted-pump approximation. The ensemble- and angle-averaged intensity $U$ of the fundamental wave (proportional to its energy density) then obeys a diffusion equation [15]

$$\begin{cases} D\dfrac{d^2 U}{dz^2} = -v I_0 \delta(z - l^*) \\ U(z = 0, z = L) = 0 \end{cases}, \qquad (21)$$

where we use the simplest boundary conditions, $D = v l^*/3$ is the diffusion constant, $v$ is the speed of wave in the average medium, $I_0$ is the intensity of the incident wave, and $l^* \approx l$ for $R \ll \lambda$. For a suspension of small spherical particles one can put $v \approx c$ as long as the suspension is not too concentrated. The solution of the above equation at $z > l^*$ is

$$U(z) = 3I_0 (1 - z/L) = 3I_0 f(z/L), \qquad (22)$$

where $f(\xi) = 1 - \xi$. For or $0 < z < l^*$, Eqs. (21) yield $U(z) = 3I_0 (z/l^*)(1 - l^*/L)$ which should be understood as an approximate result because the diffusion approximation is not expected to hold near the surface of scattering medium and source of light. For simplicity, we apply Eq. (22) for all $z$ in what follows. This does not have any considerable impact on our results as long as $L >> l^*$.

Nonlinear interaction of diffusely scattered fundamental wave with particles in suspension leads to generation of SH waves which, in their turn, are diffusely scattered by other particles. Since we assume $L >> l_2^* \approx l_2$, a typical SH wave experiences multiple scattering events before leaving the sample and hence its ensemble- and angle-averaged intensity $U_2(z)$ (proportional to the density of optical energy associated with the SH wave) can be described in the diffusion approximation. This amounts to write a diffusion equation for $U_2(z)$ with a source term $\propto U^2(z)$:

$$\begin{cases} D_2 \dfrac{d^2 U_2(z)}{dz^2} = -\beta v_2 \Sigma_2 n U^2(z) \\ U_2(z=0, z=L) = 0 \end{cases}, \qquad (23)$$

where, $D_2 = v_2 l_2^*/3$ is the diffusion constant of the SH wave, $\Sigma_2$ is given by Eq. (16), and $\beta$ is a numerical factor of order unity, describing the conversion of the ballistic SH radiation to the diffuse one at small distances from the particle. The latter conversion is a weak point of the diffusion approximation. Although it can be treated almost exactly for a plane wave incident on a slab, such a treatment appears to be rather complicated (if possible at all) in the considered case. Fortunately, the exact value of $\beta$ can only affect the absolute value of SH intensity by a numerical factor of order unity. In the following, we put $\beta = 1$.

Note that all the properties of a single particle relevant to generation of SH radiation (including parameters $\chi_1$ and $\chi_2$) enter Eq. (23) via a single parameter $\Sigma_2$ given by Eq. (16). The anisotropy of SH radiation, described by the average cosine of SH scattering angle $\langle \cos \theta \rangle$ [see Eq. (17)], has no importance in the case of concentrated suspension since anyway the radiation becomes isotropic at distances of the order of $l_2^*$ due to (linear) multiple scattering of SH waves on the particles in suspension.

The solution of Eq. (23) is readily found:

$$U_2(z) = \frac{\alpha L^2}{12}\left[1 - \frac{z}{L} - \left(1 - \frac{z}{L}\right)^4\right] = \frac{\alpha L^2}{12} f_2\left(\frac{z}{L}\right), \qquad (24)$$

where, again, we put $v_2 = c$, $\alpha = c\, \Sigma_2\, n\, (3\, I_0)^2/D_2$, and $f_2(\xi) = 1 - \xi - (1-\xi)^4$ describes the evolution of SH intensity with the depth $z$. While $f(\xi)$ decreases linearly with $z$ due to scattering, $f_2(\xi)$ shows a nonmonotonic behavior (see Fig. 4). The energy of SH wave is concentrated at a distance about $L/3$ from the front surface of the slab.

Average intensities of fundamental and SH waves at the back surface of the slab can be found by applying the Fick's law [16] that states that the flux of a quantity exhibiting diffusion behavior [in our case, optical energy associated with fundamental or SH wave and proportional to $U(z)$ or $U_2(z)$, respectively] is proportional to the gradient of its density[3]. Because intensity is proportional to the energy flux, we get

$$I \propto -\frac{D}{v}\frac{dU(z)}{dz}\bigg|_{z=L} = I_0 \frac{l^*}{L}, \qquad (25)$$

$$I_2 \propto -\frac{D_2}{v_2}\frac{dU_2(z)}{dz}\bigg|_{z=L} = \frac{3}{4} I_0 \frac{L}{\Lambda_2}, \qquad (26)$$

where $\Lambda_2$ is defined in Eq. (20). We note that the average intensity of SH wave given by Eq. (26) scales linearly with $L$. This is different from both $I_2 \propto L^2$ for SHG in a homogeneous nonlinear medium under conditions of perfect phase matching and $I_2 \propto 1/L$ for transmission of a wave through a disordered but linear medium. Curiously, the scaling $I_2 \propto L$ can be qualitatively understood by considering SHG in a disordered medium as a two-step process: first, the SH is generated in a homogeneous slab of thickness $L$, and then the resulting wave is transmitted through a disordered but linear slab of the same thickness. The second important feature of Eq. (26) is that $I_2$ does not depend explicitly on the linear scattering properties of the medium, i.e. neither on $l$ and $l_2$, nor on the corresponding transport mean free paths. It is an interesting theoretical prediction that should be possible to verify experimentally by performing SHG measurements for suspensions of particles with similar nonlinear properties (same $\Lambda_2$) but different linear scattering properties (i.e. different $l$ et $l_2$). A rather surprising feature of Eq. (26) derived in the diffusion approximation, is that it coincides (up to a numerical factor of order unity) with Eq. (19) obtained in the single scattering approximation. This suggests that the scaling law $I_2 \propto I_0 L / \Lambda_2$ can be rather general and weakly sensitive to particular experimental conditions.

---

[3] Fick's law for diffuse light is the basic relation of the diffusion approximation and can be justified both based on the phenomenological radiative transfer theory [15,16] and more rigorous diagrammatic treatment of Maxwell equations [17]. It appears to be identical to that for particle diffusion [16].

Finally, we comment on the conditions of validity of our approach. The undepleted-pump approximation requires that $I_2 \ll I$. On the other hand, the slab thickness $L$ should be large compared to $l^*$ and $l_2^*$ for the diffusion approximation to hold. We find therefore that the following inequalities should be verified for the results of the present section to be valid:

$$\frac{L}{\Lambda_2} \ll \frac{l^*}{L}, \frac{l_2^*}{L} \ll 1. \qquad (27)$$

In addition to this condition, our analysis assumes that SHG on a given particle in the bulk of the medium occurs in the field of fundamental wave resulting from incoherent sum of many partial waves scattered along various multiple-scattering paths in the medium. This implies statistical independence of these paths which is generally accepted to be true for $kl^* \gg 1$ [17]. The discussion of phase matching in the case of dilute suspension at the end of the previous section also applies (with evident modifications) in the case of concentrated suspension, requiring (a) complete disorder in particle positions and (b) large ($\gg \lambda$) inter-particle distance for our analysis to hold.

## 4. Conclusion

In the present paper we provide a theoretical description of the second harmonic generation (SHG) by a suspension of small spherical particles in the undepleted-pump approximation, assuming Raleigh limit (particle radius $R \ll \lambda$). For SHG by an isolated particle, we obtain explicit expressions for the angular diagram of SH scattering, the total power of SH wave, and the average cosine of SH scattering angle $\langle \cos\theta \rangle$. Three different contributions (dipole, quadrupole, and interference one) to the differential and total SH scattering crossections $d\Sigma_2/d\Omega$ and $\Sigma_2$, respectively, are identified and analyzed. We show that $\Sigma_2$ scales as $(R/\lambda)^6$ in contrast to the crossection $\sigma \propto (R/\lambda)^4$ of linear Raleigh scattering, and that $\langle\cos\theta\rangle$ can vary between -0.22 and 0.22 in the case of real-valued $\chi_1$ and $\chi_2$ considered in the present study. Based on the results obtained for an isolated particle, we consider SHG by (i) dilute and (ii) concentrated suspensions of identical particles, assuming that the suspension is confined within a slab of thickness $L$, illuminated by a plane monochromatic wave of intensity $I_0$. In the case (i), we apply the single scattering approximation, while in the case (ii) the diffusion approximation is used for both the fundamental and SH waves. Introducing a new SH scattering length $\Lambda_2 = 1/(\Sigma_2 n I_0)$, where $\Sigma_2$ is the SH scattering crossection of an isolated particle and $n$ is the particle number density, we show that the average intensity $I_2$ of the "transmitted" SH wave exhibits the same scaling $I_2 \propto I_0 L/\Lambda_2$ in both cases (i) and (ii). In the case (i), this is simply due to the linear increase of the total particle number with $L$, but in the case (ii) such a scaling can be understood as resulting from a combined effect of the quadratic increase of SHG efficiency with $L$, on the one hand, and attenuation of the resulting signal due to scattering, on the other hand. The angular diagram of SHG by a dilute suspension of small spherical particles shows the same qualitative features as the experimental data of Ref. 11 (in particular, SHG in the forward direction is absent). An interesting feature of the result obtained for a concentrated suspension of particles in the diffusion approximation is the independence of the SH intensity of the scattering properties of the medium for $L \gg l^*, l_2^*$. Together with the linear scaling of $I_2$ with $L$ this should be possible to verify experimentally.

**Figure captions**

**Fig. 1.**

Schematic view of the second harmonics generation (SHG) by an isolated spherical particle. Interaction of a monochromatic wave $\vec{E}_0$ (frequency $\omega$, wave vector $\vec{k}$, polarization $\hat{\varepsilon}_0$) with a small spherical particle (radius $R$) results in the linear scattering (scattered field $\vec{E}$ at frequency $\omega$) and "second harmonic" (SH) scattering (scattered field $\vec{E}_2$ at frequency $2\omega$).

**Fig. 2.**

Angular diagrams described by the functions $F_i(\theta, \varphi)$ for $i = 1$ [panel (a), dipole radiation], 2 [panel (b), quadrupole radiation), and 3 [panel (c), interference term]. The absolute value of $F_3$ is shown in the panel (c), $F_3 < 0$ for $\pi/2 < \theta < \pi$. The angular diagram resulting from a weighted sum of contributions (a-c) is shown in (d) for $\chi_1 = \chi_2$.

**Fig. 3.**

Average cosine of SH scattering angle versus $\zeta = \chi_1/\chi_2$. The maximum (minimum) of <cos $\theta$> is reached at $\zeta = \pm 2/\sqrt{5}$. Dotted lines show $\zeta = 0$ and <cos $\theta$> = 0.

**Fig. 4.**

Normalized intensities of diffuse fundamental (dashed line) and second harmonic (solid line) waves in a slab of thickness $L$ filled with a suspension of spherical particles.

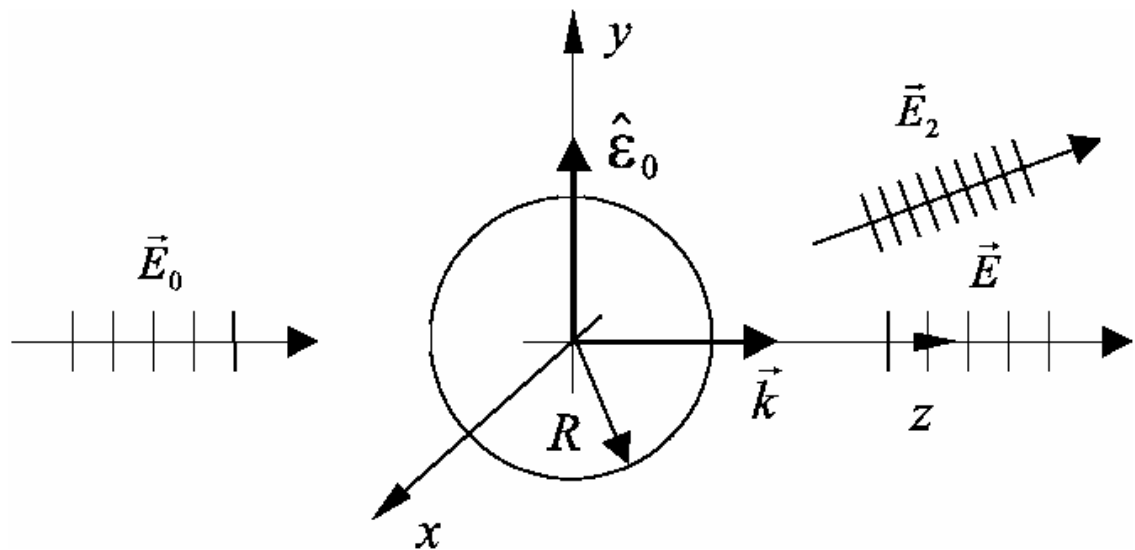

**Fig. 1**

E.V. Makeev and S.E. Skipetrov, "Second harmonic generation in suspensions of spherical particles", submitted to *Optics Communications*

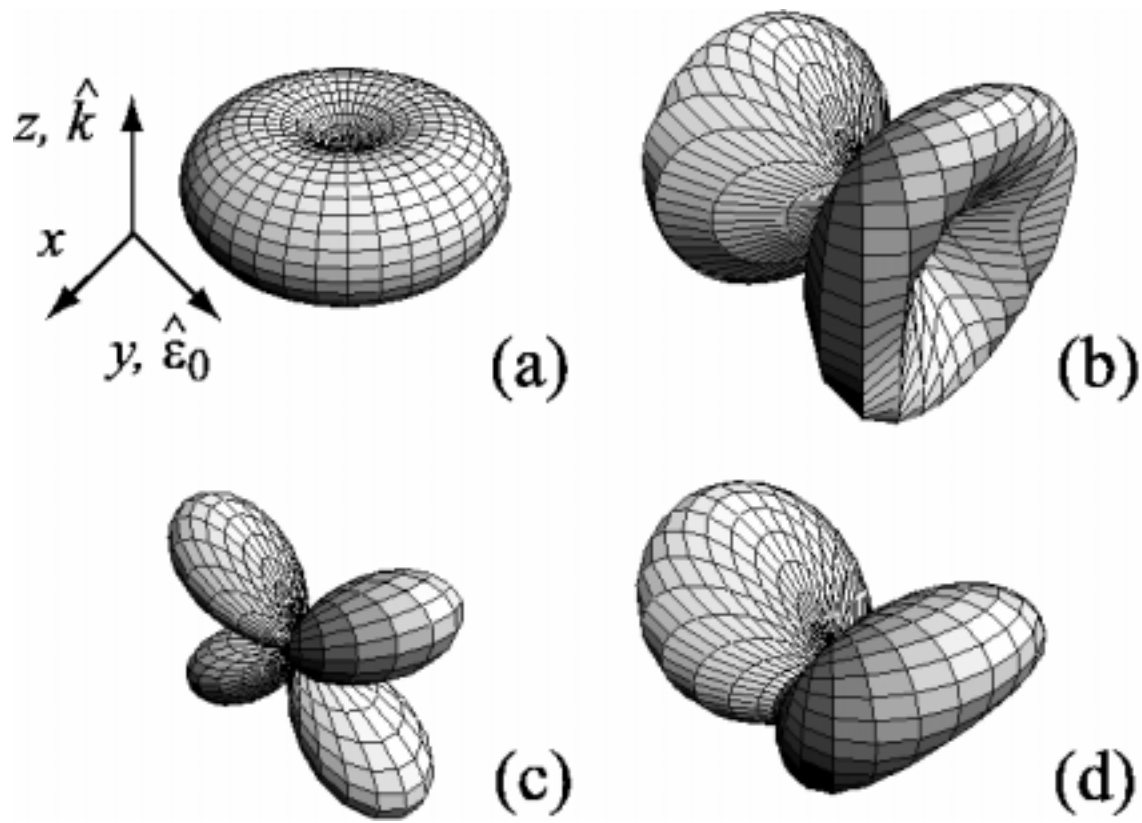

**Fig. 2**

E.V. Makeev and S.E. Skipetrov, "Second harmonic generation in suspensions of spherical particles", submitted to *Optics Communications*

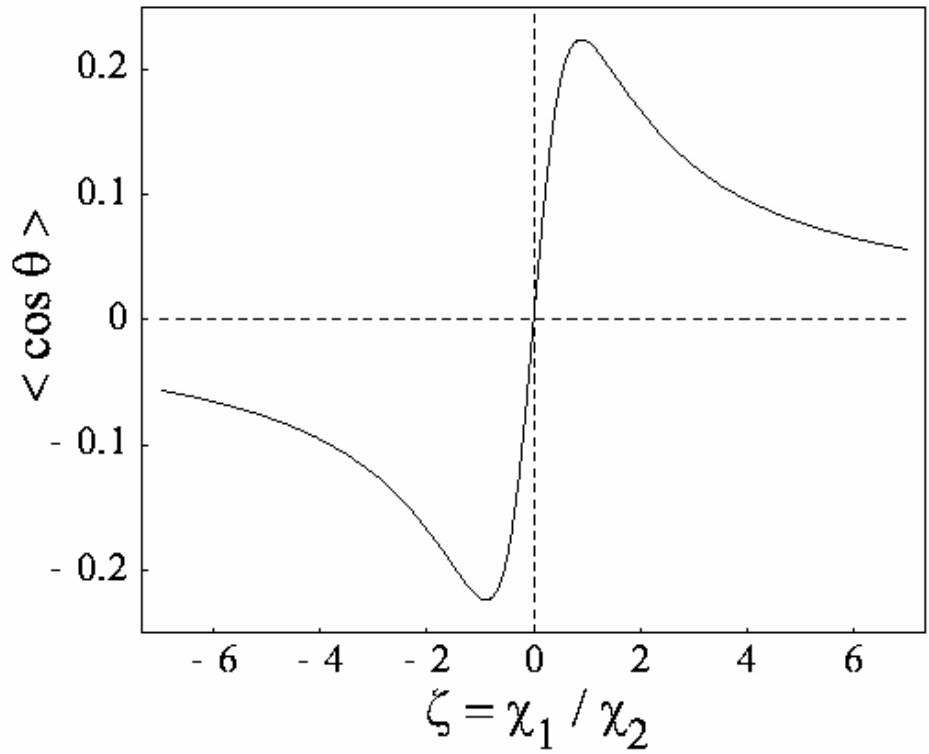

**Fig. 3**

E.V. Makeev and S.E. Skipetrov, "Second harmonic generation in suspensions of spherical particles", submitted to *Optics Communications*

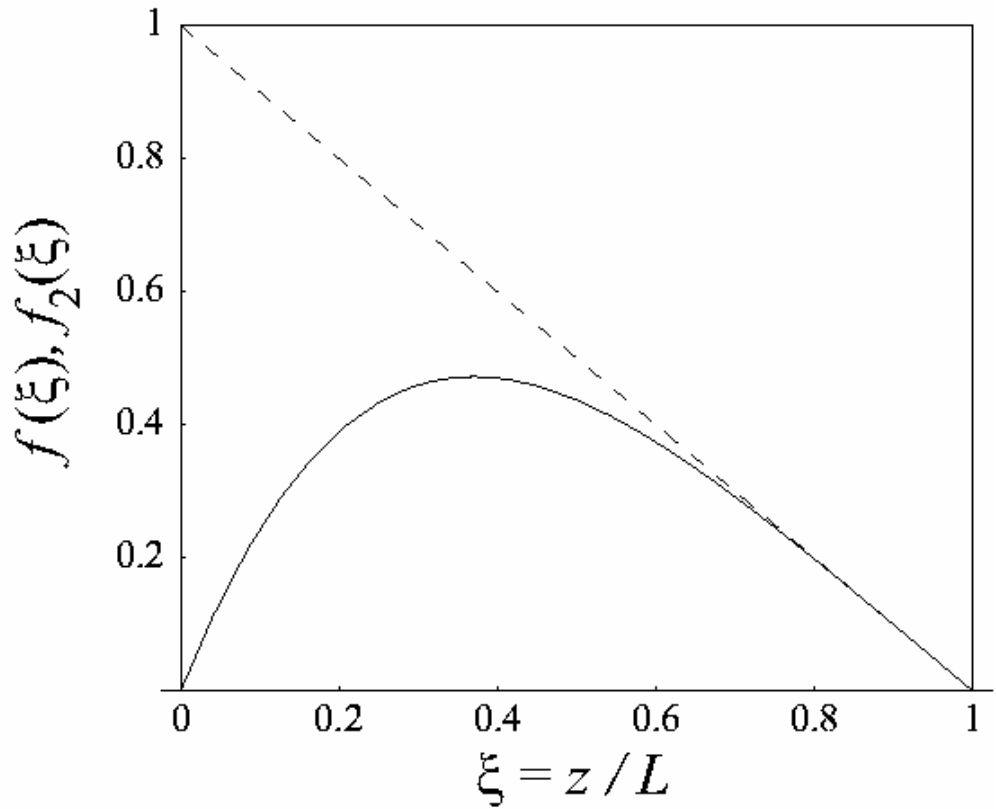

**Fig. 4**

E.V. Makeev and S.E. Skipetrov, "Second harmonic generation in suspensions of spherical particles", submitted to *Optics Communications*